# A Statistical Approach to Identifying Significant Transgenerational Methylation Changes

Ye Tian, Bai Zhang
Google Inc.
Mountain View, CA 94043, USA

Yi Fu, Guoqiang Yu, Yue Wang
Bradley Dept. of Electrical and Computer Engineering
Virginia Polytechnic Institute and State University
Arlington, VA 22203, USA

*Abstract*— Epigenetic aberrations have profound effects on phenotypic output. Genome-wide methylation alterations are inheritable to pass down the aberrations through multiple generations. We developed a statistical method, Genome-wide Identification of Significant Methylation Alteration, GISAIM, to study the significant transgenerational methylation changes. GISAIM finds the significant methylation aberrations that are inherited through multiple generations. In a concrete biological study, we investigated whether exposing pregnant rats (F0) to a high fat (HF) diet throughout pregnancy or ethinyl-estradiol (EE2)-supplemented diet during gestation days 14-20 affects carcinogen-induced mammary cancer risk in daughters (F1), granddaughters (F2) and great-granddaughters (F3). Mammary tumorigenesis was higher in daughters and granddaughters of HF rat dams, and in daughters, granddaughters and great-granddaughters of EE2 rat dams. Outcross experiments showed that increased mammary cancer risk was transmitted to HF granddaughters equally through the female or male germlines, but is only transmitted to EE2 granddaughters through the female germline. Transgenerational effect on mammary cancer risk was associated with increased expression of DNA methyltransferases, and across all three EE2 generations hypo- or hyper-methylation of the same 375 gene promoter regions in their mammary glands. Our study shows that maternal dietary estrogenic exposures during pregnancy can increase breast cancer risk in multiple generations of offspring, and the increase in risk may be inherited through non-genetic means, possibly involving DNA methylation.

## I. INTRODUCTION

Family history is a significant risk factor for breast cancer [1]. However, genetic mutations in high penetrance genes, such as BRCA1 and BRCA2, account only for a small proportion of familial breast cancers and despite intense search, no other major genetic mutations have emerged [2]. Thus, it is possible that many familial breast cancers may be transmitted not through inheritance of gene mutations, but mediated through other mechanisms, such as heritable epigenetic changes caused by in utero exposures.

Maternal exposure during pregnancy to endocrine disrupting chemicals (EDC) and dietary factors has lasting effects on breast cancer risk among female offspring. We and others have shown in animal models that exposure to compounds, such as estradiol (E2) [3], diethylstilbestrol (DES), and high fat (HF) intake during pregnancy increases daughters' (F1 generation) mammary cancer risk. Population studies also show that women who had high birth weight and daughters of women who took DES during their pregnancy are at increased breast cancer risk. Pregnant women are exposed daily to various environmental estrogenic compounds, and serum estrogens levels exhibit wide inter-individual variability among women undergoing a normal pregnancy [4].

The adverse effects of in utero exposures on adult disease are likely mediated by epigenetic dysregulation, since the epigenome is most susceptible to perturbations in early development [5]. The precise epigenetic mechanisms involved are not known but likely involve DNA methylation, as demonstrated by some studies [6]. DNA methylation is catalyzed by DNA methyltransferases (DNMT): DNMT1 maintains methylation patterns during cell division, while DNMT3a and 3b induce de novo methylation [7]. In utero exposures to endocrine disruptors alter the expression of DNA methyltransferases (DNMT1, 3a and 3b) in adult target tissues [8], but whether these changes in DNMTs also take place in the mammary gland is unknown.

More recent evidence suggests that some disease traits resulting from in utero exposures may be transmitted epigenetically through multiple generations [9]. Exposure of the developing male fetus to endocrine disruptors reduces fertility and causes prostate and kidney abnormalities that can persist for four consecutive generations [10].

In utero exposures can lead to multigenerational or transgenerational effects on disease risk. Multigenerational effects are due to direct exposure of the F1 generation embryo and F2 generation germ-line present during gestation. These effects are not transmitted to the F3 generation. In contrast, transgenerational effects are observed in at least three generations and are due to germ-line transmission since the F3 generation is not directly exposed to the environmental factor [11].

Here, we examined whether maternal exposures to HF diet or a synthetic E2 during pregnancy lead to multi or transgenerational inheritance of mammary cancer in a carcinogen-induced rat model of breast cancer. Our study shows for the first time that in utero exposures to EE2 and a HF diet can lead to trans- and multigenerational increase in the risk of breast cancer, respectively, without any further intervening exposures. The increase in breast cancer risk in the EE2 offspring is associated with heritable DNA methylation patterns in the mammary glands of all three generations studied; suggesting that the transgenerational increase in mammary cancer risk involves epigenetic inheritance. If confirmed in humans, our findings represent a novel perspective on how breast cancer risk can be transmitted across generations and could have marked implications for breast cancer prevention and treatment.

## II. METHODS

### A. Genome-wide Identification of Significant Methylation Alteration (GISAIM)

We developed a statistical approach, namely Genome-wide Identification of Significant Methylation Alteration (GISAIM), to identify the consistently inherited differential methylation patterns of genomic regions. In this study, we applied this method to identify the gene promoter regions that consistently show significant methylation changes in all three generations.

Methylation intensity fold change is used to represent the differential methylation status. The total number of bases of short reads in a promoter is used to indicate the methylation intensity of that region. Based on this measure, we calculated the fold change of methylation intensity of the $j$th promoter in the $i$th generation by

$$f_{i,j} = \frac{N_{EE2_{i,j}} + \beta L}{N_{control_{i,j}} + \beta L}, \quad i = 1,2,3 \quad (1)$$

where $N_{EE2_{i,j}}$ and $N_{control_{i,j}}$ are the methylation intensities of EE2 group and control group, respectively, $L$ is the short-read length which equals 36 in MDBCap-seq data, and $\beta$ is an offset parameter that suppresses the effects of weak methylation signals in the denominator. Here we set $\beta = 10$.

Further, we assessed the significance of the inherited differential methylation based on the methylation fold changes of all generation using GISAIM, to select promoters that have consistently large fold changes in all generations than would be expected by chance.

### B. GISAIM procedures

GISAIM follows a hypothesis testing procedure to determine if methylation changes are significant and inherited.

The summary statistics that reflecting the methylation change and inheritance is based on the methylation changes on promoter regions. We calculate the transformed fold change of all promoters by taking logarithm of the result of (1).

$$F_{i,j} = \log_2\left(\frac{N_{EE2_{i,j}} + \beta L}{N_{control_{i,j}} + \beta L}\right), \quad i = 1,2,3 \quad (2)$$

After the transformation, the fold change value become symmetric: a positive value indicates hyper-methylation and a negative value indicates hypo-methylation.

The effects of multiple generations are summed into a consistent methylation score (*M*-score) for promoter $j$ by the sum of fold changes $F_{i,j}$,

$$M_j = \sum_1^3 F_{i,j} \quad (3)$$

This definition encourages large fold changes in each and all generations and consistent changes. Contradictory fold changes will cancel each other to produce a smaller *M*-score.

Permutation test is carried out on the promoter labels within each generation and the *M*-scores for every promoter are recalculated. Suppose $B$ permutations are used, then the $B$ permuted *M*-scores of the $j$th promoter simulate the null distribution of $M_j$.

Due to the potential unbalanced occurrence between hyper-methylation and hypo-methylation, it is probable that the null distribution of $M_j$ is asymmetric in some chromosomes. Therefore, the permutation *p*-values of hyper-methylation and hypo-methylation are evaluated separately, *i.e.*, one-sided test is used in each case.

In this biological study, we find all promoters with $p<0.002$ to form significant promoter set *P*. We use *p*-value cutoff of 0.002 to select reasonable amount of promoters for visualization and further investigation.

The promoters in set *P* represent the inherited hyper-methylation and hypo-methylation induced by EE2 across three generations. An approach studying the genome-wide genetic aberrations without considering the inheritance is devised based on similar principles and has been successfully developed [12].

## III. RESULTS AND DISCUSSION

### A. Genome-wide methylation profiling

Genomic DNA (n=3/group) was isolated using the Wizard® Genomic DNA Purification Kit (Promega, Madison, WI). Methylated DNA was eluted by the MethylMiner Methylated DNA Enrichment Kit (Invitrogen). Briefly, one microgram of genomic DNA was sheared by sonication and captured by MBD proteins. The methylated DNA was eluted and used to generate Methyl-CpG binding domain-based capture (MDBCap) libraries, as previously described. MBDCap coupled with massively parallel sequencing (MBDCap-seq) libraries were sequenced using the Illumina Genome Analyzer II (GA II). Image analysis and base calling were performed with the standard Illumina pipeline. Sequencing reads were mapped by ELAND algorithm. Statistical analyses of methylation patterns across generations were focused on the gene promoter regions defined as up to 5000 base pairs upstream of transcription starting site. The CpG islands within the promoter regions were identified as previously described.

### B. Increased breast cancer risk in multiple generations of offspring

To test our hypothesis that maternal exposures during pregnancy to factors such as HF diet or a synthetic E2 lead to breast cancer in multiple generations, we fed pregnant Sprague-Dawley rats (F0) either a AIN93G control diet or an isocaloric AIN93G based HF diet, containing 43% energy from corn oil, throughout gestation. Another group of pregnant rats was fed AIN93G diet supplemented with 0.1ppm ethinyl estradiol (EE2) from day 14 to day 20 of pregnancy. F1 females were mated with F1 males from the same group to produce the F2 generation. The F3 generation was produced in the same manner. No sibling mating was carried out. The F1, F2 and F3 offspring were maintained on the AIN93G control diet for the duration of the experiment.

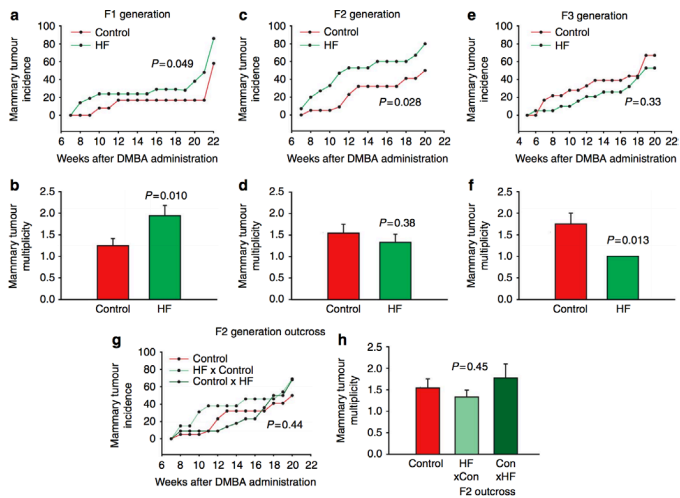

Figure 1. Multigenerational effect of maternal HF diet.

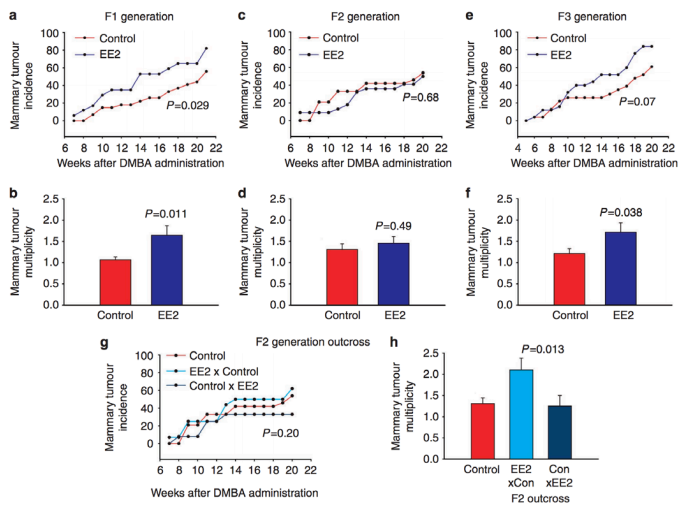

Figure 2. Transgenerational effect of maternal EE2-supplemented diet.

Effects of HF or EE2 exposure on pregnant F0 dams on mammary cancer risk in the F2 and F3 generations were then examined. In the HF granddaughters (F2 generation) mammary tumor incidence ($p=0.028$), but not multiplicity ($p=0.38$), was higher compared to the control group (Fig. 1c,d). Mammary tumor incidence did not differ ($p=0.33$) between the control and HF great-granddaughters (F3 generation), however, tumor multiplicity was lower ($p=0.013$) in the HF offspring. In the EE2 granddaughters (F2 generation), neither tumor incidence ($p=0.68$) nor multiplicity (p=0.49) was statistically different from the controls (Fig. 1e,f). However, EE2 great-granddaughters (F3 generation) had significantly higher tumor multiplicity ($p=0.038$) compared to controls (Fig. 2c,d). Mammary tumor incidence was also higher, but not statistically significant ($p=0.07$) in the EE2 F3 generation compared to controls. Histopathologic analysis indicated that the majority of mammary tumors across all three groups were malignant (adenocarcinomas or solid carcinomas).

To investigate whether mammary cancer risk could be transmitted either through the female or male lineages, we mated F1 unexposed males to F1 exposed females or F1 unexposed females to F1 unexposed males. Mammary tumor incidence was higher in both HF F2 outcross groups (HF(female)xCon(male): 68%; and Con(female)xHF(male): 69%) compared to controls (50%), but did not reach statistical significance ($p=0.44$) (Fig. 1g). These results, however, suggest that the effects of HF in utero exposure on mammary cancer risk can be transmitted through both the female or male germlines. EE2 outcross experiments show that only 33% of offspring resulting from Con(female)xEE2(male) male breeding developed mammary tumors, whilst 62% of the female EE2(female)xCon(male) offspring developed mammary tumors (Fig. 2g). Levels of tumor multiplicity were significantly higher in the EE2xCon outcross group compared to controls ($p=0.013$) (Fig. 2h). Thus, unlike HF, EE2 exposure had an opposite effect on developing male and female fetuses regarding their ability to transmit susceptibility to mammary cancer to their offspring.

### C. Altered mammary gland DNA methylation patterns in three generations of offspring.

We then determined whether the increase in Dnmt1 and Dnmt3a/3b expression in the mammary glands of EE2 offspring was associated with differential DNA methylation patterns by using Methyl-CpG Binding domain-based Capture and sequencing (MDBCap-seq) method. Using GISAIM, we examined whether any promoter regions that were differentially methylated between control and EE2 offspring in F1, F2 or F3 generations were common across all three generations. This analysis resulted in the identification of 375 differentially methylated gene promoter regions across the 21 chromosomes, among which 214 were hyper-methylated and 161 are hypo-methylated in the F1-F3 generation EE2 offspring.

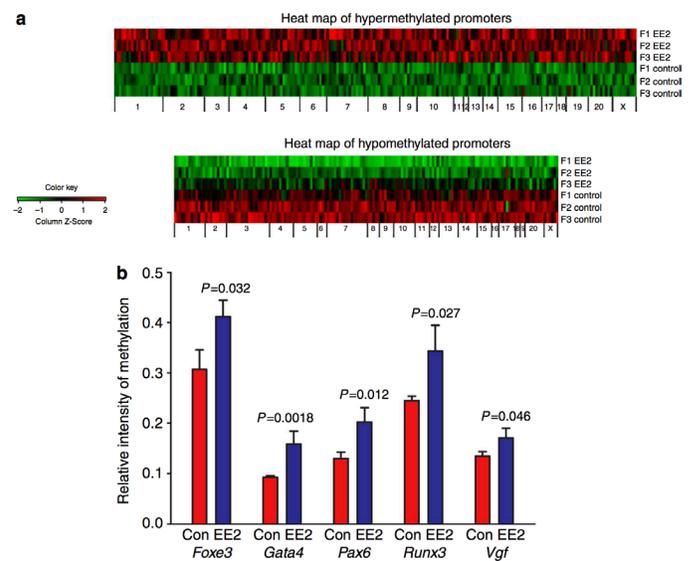

Figure 3. Differential promoter methylation in mammary glands of EE2 offspring.

Fig. 3a are the heat maps of differentially methylated gene promoter regions in EE2 offspring (green, n=3) compared with controls (red, n=3). Hyper-methylation and hypo-methylation are presented separately. mBDCap-seq method was used to access DnA methylation levels. Further analysis showed that five of the hyper-methylated promoter regions were associated with polycomb target genes (PcTGs: Pax6, Runx3, Foxe3, Gata4, Vgf) linked to stem cell differentiation and cancer. Fig. 3b shows that promoter methylation levels of the polycomb target genes (PcTGs) Foxe3, Gata4, Pax6, Runx3 and Vgf on PnD50 rat mammary glands of F1–F3 generation female offspring of Sprague-Dawley rat dams (F0) fed EE2 or control diet during gestation. Data are presented as relative intensity of methylation (mean±s.e.m., n=3 per group). $P<0.05$ is considered significant; exact *p*-values are shown in each plot.

To visually compare the inherited methylation changes of EE2 and control, we plot the methylation profiles of the 375 significant promoters of chromosomes 1 and X in Fig. 4 as examples. The chromosomes are represented by a compact view by concatenating the significant promoters while leaving remaining genomic regions out to obtain a distinguishable resolution. The horizontal axis index is base pair, and the vertical axis index is methylation count. Promoters are separated by dashed vertical lines. The start position and end position of the promoter are marked at the top-left and bottom-right of the promoter region respectively. The gene name is marked at the bottom-left of the promoter region. To facilitate the comparison, the control group is plotted under 0 by taking the opposite value. The solid blue line is the averaged methylation profile of EE2, and the solid red line is the averaged methylation profile of control. Regions in pink background indicate hyper-methylation and regions in light purple background indicate hypo-methylation. The CpG islands within the promoter regions are highlighted in the plot using dark green rectangles. We can see many methylation events are overlapped with CpG islands.

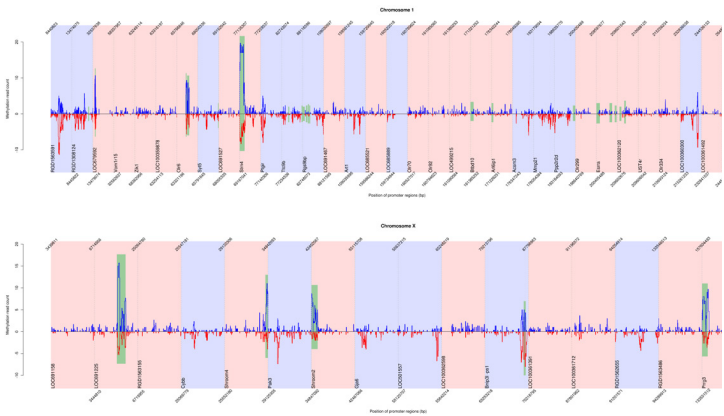

Figure 4. Inherited differentially methylated gene promoters of chromosomes 1 (upper) and X (lower).

## IV. CONCLUSION

Our statistical analyses of methylation patterns across generations are focused on the gene promoter regions defined as up to 5000 base pairs upstream of transcription starting site. To explicitly account for the background rate of random methylations and to distinguish meaningful events from random background methylations. GISAIM identifies regions of methylation change that are more likely to drive cancer risk across generations than would be expected by chance. In the breast cancer risk inheritance study, GISAIM identifies significant inheritable methylation changes between EE2+/- mice through two key steps. First, the method calculates a statistic that involves both the frequency and amplitude of the methylation change with methylation intensity being the counts of mapped methylation short reads within each of the gene prompter regions. Second, it assesses the statistical significance of each promoter methylation change using a positional permutation test that is based on the overall pattern of methylation changes seen across the gene promoter regions. GISAIM reveals a highly concordant inheritability picture involving 375 significant events ($P<0.002$), including both hyper-methylation and hypo-methylation changes in gene promoter regions. We also found many significant events (P<0.05) that are associated with well-known cancer risk genes or the CpG islands within these promoter regions.